\documentstyle[aps,eqsecnum,preprint]{revtex} \baselineskip 12pt
\begin{document}
\title{ Anti-isospectral Transformations, Orthogonal Polynomials and
Quasi-Exactly Solvable Problems. 
 }

\vspace{.4in}

\author{Avinash Khare and Bhabani Prasad Mandal}

\address{
Institute of Physics, Sachivalaya Marg,\\ Bhubaneswar-751005, India,\\
Email:  khare, bpm@iop.ren.nic.in}

\vspace{.4in}

\maketitle

\vspace{.4in}

\begin{abstract}
We consider the double
sinh-Gordon potential which is a quasi-exactly solvable problem and show
that in this case one has two sets of Bender-Dunne orthogonal polynomials .
We study in some detail the various properties of these polynomials
and the corresponding quotient polynomials. In particular, we show
that the weight functions for these polynomials are {\it not} 
always positive. We also study the orthogonal
polynomials of the double sine-Gordon potential which is related to the
double sinh-Gordon case by an anti-isospectral transformation. Finally
we discover a new quasi-exactly solvable problem by making use of the
anti-isospectral transformation. 

\end{abstract}

\newpage
\section{Introduction}

Recently, in an interesting paper, Krajewska {\it et al.} \cite{two1} have introduced
an anti-isospectral transformation ( which they also called as duality
transformation ). In particular using this transformation they  relate
the spectra of quasi-exactly solvable ( QES ) potentials $V_1$ and $V_2$
given by
\begin{eqnarray}
V_1(x) &=& x^2(ax^2+b)^2-\hbar a(2M+3)x^2 \label{66}\\
V_2(x) &=& x^2(ax^2-b)^2-\hbar a(2M+3)x^2
\label{6}
 \end{eqnarray}
where $a,b>0$ and $M$ is nonnegative integer. It may be noted here that
the duality property does not hold for other (exactly non-calculable) levels.
It is clearly of great interest to understand this new transformation 
in some detail and explore it's various consequences. For example, one
would like to know whether one can discover new QES problems by using
this transformation. Secondly are the QES levels of the dual potentials 
 related even if they  are valid over different domain? Further,  
whether the number of QES levels in  the two cases are identical
or not.

Another recent development is the work of Bender-Dunne \cite{bd}  
where they have shown that the
eigenfunctions of the Schr$\ddot{o}$dinger equation  for a quasi-exactly
solvable ( QES ) problem is the generating function for a set of orthogonal polynomials
$\{P_n(E)\}$ in the energy variable $E$. It was further shown in one specific example that these 
polynomials satisfy the three-term recursion relation 
\begin{equation}
P_n(E) = EP_{n-1}(E) +C_nP_{n-2}(E)
\label{r1}
\end{equation}
where $C_n$ is $E$ independent quantity. Using the well known theorem
\cite{the,ext},
`` the {\it necessary and sufficient} condition for a family of polynomials 
$\{P_n\}$ ( with degree $P_n= n $ ) to form an orthogonal polynomial system is 
that $\{P_n\}$ satisfy a three-term recursion relation of the form
\begin{equation}
P_n(E) = \left ( A_n E +B_n \right ) P_{n-1}(E) + C_n P_{n-2}(E), \ \ \ \ \  n\ge 1
\label{r2}
\end{equation}
where the coefficients $A_n ,\  B_n $ and $C_n$ are independent of $E$, 
$A_n \neq 0 , \ C_1 =0 ,\  C_n \neq 0 $ for $n\ge 1$", it then followed that
$\{P_n(E)\}$ for this problem forms an orthogonal set of polynomials with 
respect to some weight function, $w(E)$.
Recently several authors have studied  the Bender-Dunne polynomials 
in detail \cite{two1,ext,two,bd1}. In fact it has been claimed that the Bender-Dunne 
construction is quite universal and valid for any quasi-exactly solvable 
model in both one as well as  multi-dimensions. 

However, in a recent note \cite{qes2} we have discussed three QES problems for
all of which the Schro$\ddot{o}$dinger equation can be transformed to Heun's
equation and further in all these cases the three-term recursion relation 
satisfied by the Bender-Dunne polynomials is not of the form as given by
 Eq. (\ref{r2}) and hence does not form an orthogonal set.
It is worth pointing out that in all these cases the Hamiltonian can not be
written in terms of quadratic generator of the $Sl(2)$ algebra. We suspect
that this may be the reason why the polynomials in these examples do not
form an orthogonal set.

The question which we would like to raise and study in this paper is
regarding the properties of the Bender-Dunne polynomials in case they
form an orthogonal set. For example in the example discussed by
Bender-Dunne (as well as in most other QES examples discussed so far), for a given
potential, the QES states have either even or odd
number of nodes. The question is what happens in a QES problem if
one can obtain states with even as well as odd number of
nodes. In particular, do the polynomials corresponding to the even  and odd
number of nodes together form an orthogonal set or if the polynomials
corresponding to even number of nodes form one orthogonal set and polynomials 
with odd number of nodes form a separate orthogonal set ? Secondly, how universal
are the properties of Bender-Dunne orthogonal polynomials in such cases?
For example, are the weight functions always positive? 
Do the moments of weight function  have pure power growth? \cite{ext}
 
The purpose of this note is to explore in some detail the various issues raised
above. In particular, in Sec. II, we discuss the double sinh-Gordon (DSHG)
model which is an QES problem but where states with odd as well as even
number of nodes are known for a given potential. We show that in this case 
there are two sets of orthogonal polynomials with 
 polynomials corresponding to even number of nodes forming one orthogonal set 
while the polynomials corresponding to odd number of nodes forming another
orthogonal set. We study the properties of these polynomials as well as the corresponding
quotient  polynomials in detail and
show that in many respect they are similar to the Bender-Dunne polynomials.
However, unlike the Bender-Dunne case, the weight functions are not always positive for either of the
orthogonal polynomial sets.
In Sec. III, we discuss several consequences of the duality symmetry in the
context of the QES problems. In particular, we study two QES problems, in
both of which the domain of validity  of the potential and it's dual are different.
As a result we show that in the case of the double sine-Gordon ( DSG )
potential, the number of QES levels
are almost half of those in the double sinh-Gordon case ( which is dual to it ). 
However, even then many of the predictions about their corresponding
 Bender-Dunne polynomials  still go through. For example,
the QES levels of DSG are still related to the corresponding levels of
DSGH case. Further, the weight functions
here are also not always positive. 
Finally using duality, we discover a new QES problem
which has so far not been discussed in the literature. Sec. IV is reserved
for discussions.

\section{ DSHG system and Orthogonal Polynomials}

The DSHG system is one of the few double well problems in quantum mechanics which
is quasi-exactly solvable \cite{raz,hks}. This double well system has
found application in several different branches of physics
starting from the theory of diffusion in bistable field \cite{bis}
, quantum theory of instantons \cite{pkov,lee}, quantum theory of molecules 
and also as a model for nonlinear coherent structure \cite{khs,snb}. The Hamiltonian
for this case is given by ( we shall assume $\hbar = 2m =1$ throughout this paper) 
\begin{equation}
H= - \frac{ d^2}{dx^2} + \left ( \zeta\cosh 2x -M \right )^2
\label{h}
\end{equation}
where $\zeta$ is a positive parameter. The value of $M$
is not restricted in principle but it has been shown \cite{raz} that the
solutions for first $M$ levels are exactly known in case $M$ is a positive
integer. Note further that for $M>\zeta$, this potential has a double well
structure with the two minima lying at $\cosh 2x_0 = \frac{ M}{\zeta}$.
Let us now derive the three-term recursion relation in this case. On  
substituting
\begin{equation}
\psi(x) = e^{-\frac{ \zeta}{2}\cosh 2x}\phi(x)
\end{equation}
in the Schro$\ddot{o}$dinger equation $H\psi= E\psi$ with $H$ as given by
Eq. (\ref{h}) we obtain
\begin{equation}
\phi^{\prime\prime}(x) -2\zeta\sinh 2x\phi^\prime (x) + \left
 [(E-M^2-\zeta^2) +2(M-1)\zeta\cosh 2x \right ]\phi(x) =0
\end{equation}
On further substituting
\begin{equation}
z=\cosh 2x -1;\ \ \ \phi = z^s\sum_{n=0}^\infty \frac{ R_n(E)}{n!} 
\left ( \frac{ z+2}{2} \right )^{ \frac{ n}{2}}
\label{5}
\end{equation}
we obtain the three-term recursion relation $(n\ge 0)$.
\begin{eqnarray} 
R_{n+2}(E)&-&\left [ n^2 +4(s+\zeta)n + 4s^2 + \left .
E-M^2-\zeta^2-2(M-1)\zeta \right .  \right ]R_n(E) \nonumber \\  
-&&4\zeta \left [ M+1-2s-n \right ]n(n-1) R_{n-2}(E)=0
\label{r22}
\end{eqnarray}
provided $2s^2 = s$ i.e. either $s=0$ or
$s= \frac{ 1}{2}$. Thus we have two sets of independent solutions 
: the even states (i.e. states with even number of nodes) for $s=0$
and the odd states for $s= \frac{ 1}{2} $. Note that unlike the
Bender-Dunne (or most other QES) cases, $s$
is not contained in the potential and this is perhaps related to 
the fact that for any integer $M \ (\ge 2)\ $ the QES 
solutions corresponding to both even and odd states are obtained.
From Eq. (\ref{r22}) we observe that the even and odd
polynomials $R_n(E)$ do not mix with each other and hence 
we have two separate three-term recursion relations
depending on whether $n$ is odd or even. In particular, it is easily 
shown that the three-term recursion relations corresponding to the even and odd 
$n$ cases respectively are given by $(n\ge 1)$
\begin{eqnarray}
P_n(E)&-& \left [ 4n^2 +8n(s+\zeta-1)+4s^2-8s +4-6\zeta +E -(M+\zeta)^2
\right ]P_{n-1}(E) \nonumber \\   
&-& 8\zeta(n-1)(2n-3)\left ( M+3-2s-2n \right ) P_{n-2} = 0 \label{rr0}  \\  
Q_n(E)&-& \left [ 4n^2 +4n(2s+2\zeta-1)+4s^2-4s +1-2\zeta +E -(M+\zeta)^2
\right ]Q_{n-1}(E) \nonumber \\  
&-& 8\zeta(n-1)(2n-1)\left ( M+2-2s-2n \right ) Q_{n-2} = 0   
\label{rr1}
\end{eqnarray}
with $P_0(E)=1, Q_0(E)=1. $ 
These recursion relations generate a set of monic polynomials, of which
 the first few  are  
\begin{eqnarray} 
P_0(E) &=&1 \nonumber \\  
P_1(E) &=& {\cal E}+4s^2+2\zeta \nonumber \\ 
P_2(E)&=& {\cal E}^2 +{\cal E} \left [ 8s^2+8s+ 12\zeta +4 \right ]
\nonumber \\ &+& \left ( 4s^2+4s+10\zeta +4 \right ) \left ( 4s^2+2\zeta \right )
+ 8\zeta  \left (M-2s-1 \right )
\end{eqnarray}
and
\begin{eqnarray} 
Q_0(E) &=&1 \nonumber \\  
Q_1(E) &=& {\cal E}+4s^2+4s +1+ 6\zeta \nonumber \\  
Q_2(E)&=& {\cal E}^2 +{\cal E} \left [ 8s^2+8s+ 20\zeta +10\right ]
\nonumber \\ &+& \left ( 4s^2+4s+6\zeta +1 \right ) \left ( 4s^2+4s+14\zeta +9\right )
+24\zeta  \left (M-2s-2 \right )
\end{eqnarray}
 where 
\begin{equation}
{\cal E} \equiv E-(M+\zeta)^2
\label{e}
\end{equation}
Note that unlike the Bender-Dunne case, the polynomails $P_n(E)$ and $Q_n(E)$
are not eigenfunctions of parity.
It is easily seen that when $M$ is a positive integer, exact solutions for
first $M$ levels are obtained. In particular, if $M$ is odd (even) integer, then
solutions with even number of nodes $(s=0)$ are obtained when the
coefficient of $P_{n-2}\  (Q_{n-2})$ vanishes. Similarly if $M$
is odd (even) integer, the solutions with odd number of nodes $(s =\frac{ 1}{2} )$ are
obtained when the coefficient of $Q_{n-2}\ (P_{n-2})$ vanishes. Further
for $M$ even (say $2k+2,\ \  k=0,1,2\cdots$ ), half the levels, i.e. $(k+1)$ levels are obtained each from the zeros of the
orthogonal polynomials $P_{k+1}(E)$ and $Q_{k+1}(E)$. On the other hand 
when $M$ is odd (say $2k+1, \ \ \ k=0,1\cdots)$ then $k+1$ and $k$ levels are obtained  from the  zeros
of the orthogonal polynomials $P_{k+1}(E)$ and $Q_{k}(E)$ respectively.

We have studied the properties of the polynomial sets \{$P_n(E)$\} and \{ $Q_n(E)$\}
and we find that many of their properties are almost the same and in turn they are 
similar to the Bender-Dunne polynomials . In particular, since in both the
cases
 the recursion relations are similar to those given by  Eq. (\ref{r2}) , hence
for
all values of the parameters $M$ and $s$, they form an orthogonal set .
Secondly, the wave function $\psi(x,E)$ is the generating function for the
polynomials $P_n(E)$ as well as $Q_n(E)$. Thirdly when $M$ is a positive
integer, both of these polynomials  exhibit factorization property whose precise form
depends on whether $M $ is even or odd.
In particular, if $M=2k+1, \ k=0,1,2\cdots $ then one has $(n\ge 0)$
\begin{eqnarray}
P_{k+1+n}(E)&=& P_{k+1}(E)\bar{P}_n(E) \nonumber \\ 
Q_{k+n}(E)&=& Q_{k}(E)\bar{Q}_n(E) 
\label{f1}
\end{eqnarray}
On the other hand if $M= 2k+2,\ k=0,1,2\cdots $, then 
\begin{eqnarray}
P_{k+1+n}(E)&=& P_{k+1}(E)\bar{R}_n(E) \nonumber \\ 
Q_{k+1+n}(E)&=& Q_{k+1}(E)\bar{S}_n(E) 
\label{f2}
\end{eqnarray}
where $\bar{P}_0(E)=\bar{Q}_0(E) =\bar{R}_0(E)=\bar{S}_0(E)=1.$
Following Ref. \cite{two}  it is easily shown that the polynomial sets $\{\bar{P}_n(E)\}$, 
$\{\bar{Q}_n(E)\}$ , $\{\bar{R}_n(E)\}$   and $\{\bar{S}_n(E)\}$ 
correspond to the non-exact spectrum for this problem.

To illustrate this factorization we list in factored form,
the first few polynomials of both the types in case $M=3 $ and $M=4$.
\begin{eqnarray}
\underline{M=3}&& \nonumber \\ 
\nonumber \\ 
P_0(E)&=&1 \nonumber \\ 
P_1(E) &=& {\cal E} +2\zeta \nonumber \\ 
P_2(E) &=& {\cal E}^2 +{\cal E} \left [ 12\zeta +4 \right ] +20\zeta^2
+24\zeta \nonumber \\ 
P_3(E) &=& \left [ {\cal E} + 18\zeta +16 \right ]P_2(E) \nonumber \\ 
P_4(E) &=& \left [ {\cal E}^2 + {\cal E} (46\zeta +52) +(36
+28\zeta)(16+18\zeta)-240\zeta \right ]P_2(E).
\label{mp3}
\end{eqnarray}
\begin{eqnarray}
Q_0(E)&=&1 \nonumber \\ 
Q_1(E) &=& {\cal E} +6\zeta+4 \nonumber \\ 
Q_2(E) &=& \left [{\cal E} + 14\zeta +\frac{ 35}{2} \right ]Q_1(E)
 \nonumber \\ 
Q_3(E) &=& \left [ {\cal E}^2 + {\cal E} (36\zeta +52) +(36
+22\zeta)(16+14\zeta)-160\zeta \right ]Q_1(E).
\label{mq3}
\end{eqnarray}
where ${\cal E} $ is given by Eq. (\ref{rr0}).
Let us notice that in this case $P_2(E)$, is a common factor of $P_n(E)$ 
for $n\ge2$. The zeros of $P_2(E)$ are at
\begin{eqnarray}
E_0 &=& 7+\zeta^2- 2 \sqrt{1+4\zeta^2} \nonumber \\  
E_2 &=& 7+\zeta^2+ 2 \sqrt{1+4\zeta^2}   
\label{02}
\end{eqnarray} 
which give the ground and second excited state eigenvalues for the
potential when $M =3$. On the other hand  $Q_1(E)$ is a common
factor of $Q_n(E)$ for $n\ge 1$ . The zeros of $Q_1(E)$ are
at 
\begin{equation}
E_1= \zeta^2 +5
\end{equation}

which give the first excited state eigenvalue for the same potential
(i.e. when $M=3$). In this way for $M=3$ one obtains exact energy
eigenvalues for the first three levels.

\begin{eqnarray}
\underline{M=4}&& \nonumber \\ 
\nonumber \\ 
P_0(E)&=&1 \nonumber \\ 
P_1(E) &=& {\cal E} +2\zeta +1\nonumber \\ 
P_2(E) &=& {\cal E}^2 +{\cal E} \left [ 12\zeta +10\right ] +20\zeta^2
+44\zeta +9 \nonumber \\ 
P_3(E) &=& \left [ {\cal E} + 18\zeta +25 \right ]P_2(E) \nonumber \\ 
P_4(E) &=& \left [ {\cal E}^2 + {\cal E} (46\zeta +64) +(
28\zeta+49)(18\zeta+25)-240\zeta \right ]P_2(E).
\label{mp4}
\end{eqnarray}
\begin{eqnarray}
Q_0(E)&=&1 \nonumber \\ 
Q_1(E) &=& {\cal E} +6\zeta+1 \nonumber \\ 
Q_2(E) &=& {\cal E}^2 +{\cal E} \left [ 20\zeta +10\right ] +84\zeta^2
+116\zeta +9 \nonumber \\ 
Q_3(E) &=& \left [{\cal E} + 22\zeta + 25 \right ]Q_2(E)
 \nonumber \\ 
Q_4(E) &=& \left [ {\cal E}^2 + {\cal E} (52\zeta +74) +(30\zeta+49)
(22\zeta+25)-336\zeta \right ]Q_2(E).
\label{mq4}
\end{eqnarray}
In this case $P_2\ (Q_2)$ is a common factor of $P_n\ (Q_n)$ for $n\ge 2$
. The zeros of $P_2(E)$ are at
\begin{eqnarray} 
E_1& =& \zeta^2 +2\zeta+11-4\sqrt{\zeta^2+\zeta+1} \nonumber \\  
E_3 &=& \zeta^2 +2\zeta+11+4\sqrt{\zeta^2+\zeta+1}
\label{13}
\end{eqnarray}
which give the first  and the third excited state energies for the
potential when $M=4$. On the other hand
the zeros of $Q_2(E) $ are at 
\begin{eqnarray} 
E_0& =& \zeta^2 -2\zeta+11-4\sqrt{\zeta^2-\zeta+1} \nonumber \\  
E_2 &=& \zeta^2 -2\zeta+11+4\sqrt{\zeta^2-\zeta+1}
\label{32}
\end{eqnarray}
which give the ground  and the second  excited state energies for the same potential $(M=4)$.
The corresponding eigenfunctions can be easily obtained by evaluating
$\psi(x)$
in Eq. (\ref{5}) at these values of $E$ and not surprisingly they are the same
as given by Razavy \cite{raz} ( with appropriate change of parameters
).

The norms (squared) of the orthogonal set of polynomials $P_n(E)$ and $Q_n(E)$ can be
easily determined by using the recursion relations and we find
( assuming $\gamma _0^{(p)}= \gamma _0^{(q)}=1$)
\begin{eqnarray}
\gamma _n^{(p)} &=& (-8\zeta)^n n! \prod_{k=1}^n \left [(2k-1) \left   
(M-2s-2k+1 \right ) \right ]\nonumber \\ 
\gamma _n^{(q)} &=& (-8\zeta)^n n! \prod_{k=1}^n \left [(2k+1)\left   
(M-2s-2k \right ) \right ] 
\label{24}
\end{eqnarray}
We observe from here that $\gamma _n^{(p)}$  vanishes
for $2n\ge M-2s+1 $ while $\gamma _n^{(q)}$ vanishes for $2n\ge M-2s$
provided $M$ is a positive integer and as remarked by Bender-Dunne, this
vanishing of norm  is an alternative characterization of the QES problem.
It may be noted that unlike the Bender-Dunne case, the norms are
alternative in sign for $2n <M-2s+1$ or $2n<M-2s$ depending on the set
$\{P_n(E)\}$ or $\{Q_n(E)\}$ respectively. 

Using the factorization property as given by Eqs. (\ref{f1}) and (\ref{f2})
 and using the recursion relations (\ref{rr0}) and (\ref{rr1})
satisfied by $P_n$ and $Q_n$, it is easily shown that $\bar{P}_n\ \bar{Q}_n \ 
\bar{R}_n \ $ and $ \bar{S}_n $   
satisfy the following recursion relations
\begin{eqnarray}
\bar{P}_n(E) &=& \left [ (M+1+2n)^2 +4(M+1 +2n)(s+\zeta-1) +
4s^2-8s+4-6\zeta +{\cal E} \right ] \bar{P}_{n-1}(E) \nonumber \\  
&+& 4\zeta(M-1+2n)(M+2n-2) \left [2-2s-2n \right ]\bar{P}_{n-2}(E) 
\nonumber \\  
\bar{Q}_n(E) &=& \left [ (M-1+2n)^2 +2(M-1 +2n)(2s+2\zeta-1) +
4s^2-4s+1-2\zeta +{\cal E} \right ] \bar{Q}_{n-1}(E) \nonumber \\  
&+& 4\zeta(M-3+2n)(M+2n-2) \left [3-2s-2n \right ]\bar{Q}_{n-2}(E) 
\nonumber \\ 
\bar{R}_n(E) &=& \left [ (M+2n)^2 +4(M +2n)(s+\zeta-1) +
4s^2-8s+4-6\zeta +{\cal E} \right ] \bar{R}_{n-1}(E) \nonumber \\  
&+& 4\zeta(M-1+2n)(M+2n-3) \left [3-2s-2n \right ]\bar{R}_{n-2}(E) 
\nonumber \\ 
\bar{S}_n(E) &=& \left [ (M+2n)^2 +2(M +2n)(2s+2\zeta-1) +
4s^2-4s+1-2\zeta +{\cal E} \right ] \bar{S}_{n-1}(E) \nonumber \\  
&+& 4\zeta(M-2+2n)(M+2n-1) \left [2-2s-2n \right ]\bar{S}_{n-2}(E) 
\label{25}
\end{eqnarray}

Using these recursion relations, it is straightforward to obtain the norms
of theses polynomials and show that they are all positive. In particular,
we obtain 
\begin{eqnarray}
\gamma _n^{\bar{P}}&=& (4\zeta)^n\prod_{k=1}^n \left [ (M+2k+1)(M+2k)
(2k+2s) \right ]\nonumber \\  
\gamma _n^{\bar{Q}}&=& (4\zeta)^n\prod_{k=1}^n \left [ (M+2k)(M+2k-1)
(2k+2s-1) \right ]\nonumber \\  
\gamma _n^{\bar{R}}&=& (4\zeta)^n\prod_{k=1}^n \left [ (M+2k+1)(M+2k-1)
(2k+2s-1) \right ]\nonumber \\  
\gamma _n^{\bar{S}}&=& (4\zeta)^n\prod_{k=1}^n \left [ (M+2k+1)(M+2k)
(2k+2s) \right ]\nonumber \\  
\label{4n}
\end{eqnarray}
 Note that in Eqs. (\ref{25}) and (\ref{4n}) $M$ and $s$ take only specific
values. For example, in the case of $\bar{P}_n(E)\ \  M$ is odd and $s=0$,
while in the case of $\bar{Q}_n(E) \ \ M$ is odd and $s=\frac{1}{2}$.
On the other hand, in the case of $\bar{R}_n(E)\ \ M$ is even and $s=\frac{1}{2}$
while in the case of $\bar{S}_n(E)\ \ M$ is even but $s=0$.

Finally, we can also obtain the weight function $w(E)$ for our polynomials 
$\{P_n(E)\}$ and $\{Q_n(E)\}$ by using the expression as derived by  Krajewska {\it et al} \cite{two}.  
In particular, the weight function $w^{(p)}(E)$ can be written as
\begin{equation}
w^{(p)}(E) = \sum_{k=1}^M w^{(p)}_k \delta(E-E_k)
\label{ww}
\end{equation}
where the numbers $w^{(p)}_k$ satisfy the algebraic equation
\begin{equation}
\sum_{k=1}^M P_n(E_k)w^{(p)}_k = \delta _{n0}
\label{ww1}
\end{equation}
A similar relation also exists for the weight function of the  $Q$ polynomials. 

As an illustration we have computed the weight functions $w_n^{(p)}$ and
$w_n^{(q)}$ in case $M=3$ and 4. For example, when $M=3$, using Eqs.
(\ref{ww}) to (\ref{ww1}) and (\ref{mp3}) we find that
\begin{eqnarray}
w_0^{(p)} &=& \frac{ 1}{2} -\frac{ 2\zeta+1 }{2\sqrt{1+4\zeta^2}} \nonumber \\  
w_2^{(p)} &=& \frac{ 1}{2} +\frac{ 2\zeta+1 }{2\sqrt{1+4\zeta^2}} \nonumber \\  
w_1^{(q)} &=& 1
\label{wm3}
\end{eqnarray}
On the other hand, when $M=4$, using Eqs. (\ref{ww}),
(\ref{ww1}),(\ref{mp4}) and (\ref{mq4})
we find 
\begin{eqnarray}
w_0^{(q)} &=& \frac{ 1}{2} - \frac{ \zeta+1}{2\sqrt{\zeta^2-\zeta+1}} \nonumber \\ 
w_1^{(p)} &=& \frac{ 1}{2} - \frac{ \zeta+1}{2\sqrt{\zeta^2+\zeta+1}} \nonumber \\ 
w_2^{(q)} &=& \frac{ 1}{2} + \frac{ \zeta+1}{2\sqrt{\zeta^2-\zeta+1}} \nonumber \\ 
w_3^{(p)} &=& \frac{ 1}{2} + \frac{ \zeta+1}{2\sqrt{\zeta^2+\zeta+1}} 
 \label{wm4}
\end{eqnarray} 
As a cross check on our calculation of weight functions, we have calculated
the norm of the polynomials by using the basic relations
\begin{equation}
\gamma _n^{(p)} = \int dE w^{(p)}(E)[P_n(E)]^2; \ \ \ \ 
\gamma _n^{(q)} = \int dE w^{(q)}(E)[Q_n(E)]^2
\end{equation}
and for $M=3$ as well as 4 we have verified that we get the same answer as
given by Eq. (\ref{24}).
It is worth pointing out that unlike the Bender-Dunne example, in 
none of our cases the weight functions are always positive. Actually this is
not all that surprising. As has been proved by Finkel {\it et al.} \cite{ext},
if the three-term recursion relation is of the form
\begin{equation}
\hat{P}_{k+1} = (E-b_k)\hat{P}_k -a_k\hat{P}_{k-1}, \ \ \ k\ge 0
\label{fin}
\end{equation}
with $a_0=0$ and $a_{n+1}=0$, then the weight functions are all positive if
$b_k$ is real for $0\le k\le n$ and $a_k >0$ for $1\le k\le n$.
On comparing Eq. (\ref{fin}) with our recursion relation (\ref{rr0}) and
(\ref{rr1})
we find that in both of our cases, $a_k$ is in fact $<0$ for $1\le k\le n$.

Once the weight function is known, then one can also calculate the moments of
$w(E)$, defined by  
\begin{equation}
\mu_n = \int dE w(E) E^{n}
\end{equation}
Since in our case, the polynomials $\{P_n(E)\}$ and $\{Q_n(E)\}$ are not eigenfunctions
of parity, hence unlike the Bender-Dunne case, odd moments do not vanish in our
case. For example, the first few moments in our case for $M=3$ are ( we choose $\mu_0^{(p)}=
\mu_0^{(q)}=1$ without any loss of generality )
\begin{eqnarray}
\mu_1^{(p)} &=& (3+\zeta)^2 -2\zeta ; \ \ \ \  \mu_2^{(p)} =-16\zeta +[(3+
\zeta)^2 -2\zeta]^2 \nonumber \\  
\mu_1^{(q)} &=& (3+\zeta)^2-6\zeta -4;\  \ \ \  \mu_2^{(q)} = [(3+
\zeta)^2 -6\zeta -4]^2 
\end{eqnarray}
Following Finkel {\it et al. }\cite{ext}, we can also calculate the growth
rate of the moments. In particular, since in both of our cases, $a_k\neq 0 $ for
$1\leq k<n $ where $a_k$ as given by Eqs. (\ref{fin}), (\ref{rr0}) and \ref{rr1},
hence following their discussion it is easily shown that for large $n$, to leading
order
\begin{equation}
\mu_n^{(p)},\ \ \mu_n^{(q)} \sim (M+\zeta)^{2n}
\end{equation}
Thus in our case the moments  have a pure power growth.

\section{ New QES Potentials From Anti-isospectral Transformations }

In an interesting paper, Krajewska {\it et al}. \cite{two1} have recently discussed
the consequences of the anti-isospectral transformation ( also termed
as duality transformation ) in the context of the QES problems. In particular,
they have shown that under the transformations $x\rightarrow ix=y $, if  a potential 
$V(x)$ goes to $\bar{V}(y)$, i.e.
\begin{equation}
V(x)\longrightarrow \bar{V}(y)
\end{equation}
then  the QES levels of the two  are also related. In particular, they have 
shown that if $M$ levels of the potential $V(x)$ are QES levels  with
energy eigenvalues and eigenfunctions $E_k(k=0,1, \cdots, M-1)$ and $\psi_k(x)$
respectively then the energy eigenfunctions of $\bar{V}(y)$ are given by
\begin{equation}
\bar{E}_k = -E_{M-1-k}, \ \ \ \bar{\psi}_k(y) = \psi_{M-1-k}(ix)
\label{35}
\end{equation}

As an illustration, these authors have discussed  the $x^6$- potentials
as given by Eqs. (\ref{66}) and (\ref{6}) and explained in details how the eigen 
spectra of these two dual potentials are related. However in this particular
case, the domain of $x$ and $y$ are the same.

The purpose of this section is to explore the consequences of this symmetry
when the domain of validity of $V(x)$ and $\bar{V}(y)$ are different and 
to see how many of the results derived in Ref. \cite{two1} go through in this case.
In particular, do $V(x)$ and $\bar{V}(y)$, hold the same number of bound states
? Are the bound states of the two potentials still related 
by Eq. (\ref{35})? 
 To begin with, we would  also like to comment that, the duality symmetry is only useful
in case the potential $V(x)$ is symmetric in $x$ as otherwise the dual potential 
$\bar{V}(y)$ will be a complex potential .
 
\subsection{DSG Example}

Let us consider the potential corresponding to the DSHG case, and
try to explore the consequences of the duality symmetry in this case. On
applying the duality transformation $x\rightarrow i \theta $ to the
Schro$\ddot{o}$dinger equation $ H\psi = E\psi $ with $H$ as given by
Eq. (\ref{h}) we obtain the following Schr$\ddot{o}$dinger equation for
the DSG equation.
\begin{equation}
\left [-\frac{ d^2}{d \theta ^2}-\left (\zeta\cos2 \theta -M \right )^2
\right ]\psi(\theta )= \hat{E}\psi( \theta ), \ \ \  \hat{E}= -E.
\end{equation}
Note that whereas the domain of $x$ in DSHG equation is $-\infty \le x\le
\infty$
in the DSG case, the domain is $0\le \theta \le 2\pi$ provided we are solving the
problem on a circle as we do here. As a result,
unlike the DSHG case, in the DSG case  due to the
periodicity constraint, one has to demand that the  wave function $\psi(\theta
)$ must be invariant under $\theta \rightarrow \theta + \pi$ i.e.
\begin{equation}
\psi(\theta +\pi) = \psi( \theta )
\end{equation}
Looking
at the QES eigenfunctions of the DSHG case, it is easy to see that when
$M$ is even $(i.e.\ M =2,4 \cdots)$ then the eigenfunctions of the DSG case do not remain invariant
under $\theta \rightarrow \theta +\pi$ but they change sign. As a result,
the QES levels of DSG when $M$ is even must be rejected.
Thus on physical ground, we find that in case the domain of
$V(x)$ and $\bar{V}(y)$ are different then the number of QES levels of
$V(x)$ and $\bar{V}(y)$ need not be identical. In particular, boundary conditions
may forbid some wave functions to be eigenfunctions. For example,
for $M=2$, naively one would have two exact energy eigenstates of DSG
equation as given by
\begin{eqnarray} 
\psi^{DSG}_0(\theta )&=& \sin \theta e^{-\frac{ \zeta}{2}\cos 2 \theta
}\nonumber \\  
\psi^{DSG}_2(\theta )&=& \cos \theta e^{-\frac{ \zeta}{2}\cos 2 \theta }
\end{eqnarray} 
However both of these are unacceptable eigenfunctions of DSG, being odd under $\theta 
\rightarrow \theta +\pi$. Thus we have shown that even though DSG and DSHG
are dual system, DSG equation has approximately only half the number of QES levels compared
to the DSHG case. In particular, when $M$ is an odd integer $( M= 1,3,5\cdots)$
 only then the DSG is a QES system and in that case the first $M$ levels are exactly
known, and in fact they can be immediately obtained from the corresponding
DSHG case by making use of the duality relation (\ref{35}).
 For example, for $M=1$ the ground state of the above DSG is given by
\begin{equation}
\hat{E}_0= -(1+\zeta^2)  \ \ \ \ \ \ \ \psi_0 = e^{-\frac{ \zeta}{2}\cos 2 \theta
}
\end{equation}
On the other hand for $M=3$, the first 3 levels of the DSG equation are
are given by
\begin{eqnarray}
\hat{E}_0 &=& -7-\zeta^2-2\sqrt{1+4\zeta^2}; \ \ \ \   \psi_0= \left [ 2\zeta-
\left \{ \sqrt{1+4\zeta^2}-1 \right \}\cos 2 \theta \right ]e^{-\frac{\zeta}{2}
\cos 2 \theta } \nonumber \\
\hat{E}_1 &=& -\zeta^2 -5;\ \ \ \ \ \ \ \  \ \ \ \ \ \ \ \ \ \ \psi_1=\sin 2 \theta e^{-\frac{ \zeta}{2}
\cos 2 \theta } \nonumber \\   
\hat{E}_2 &=& -7-\zeta^2+2\sqrt{1+4\zeta^2};\ \ \ \  \psi_2= \left [ 2\zeta+
\left \{ \sqrt{1+4\zeta^2}+1 \right \}\cos 2 \theta \right ]e^{-\frac{\zeta}{2}
\cos 2 \theta } 
\label{123} 
\end{eqnarray}
The energy eigenstates of the DSG equation have recently been obtained
by Habib {\it et al.}\cite{hks} and not surprisingly our results are the same
as obtained by them.

One can now study the Bender-Dunne polynomials of the DSG equation and it
is easy to see that most of the discussion of the last section ( for the DSHG case ) 
 also goes through in this case except that now only odd values of
$M$ give us a QES system. For example, the recursion relations for $P_n$ and $Q_n
$ as given in the last section are also valid in this case provided we replace
$E$ by $-E$. Thus the three-term recursion relation
for $P_n$ as given by Eq. (\ref{rr0}) with $E$ changed to $-E$ will 
give an exact solution for the DSG case only if $s=0$ while the three-term recursion relation 
for $Q_n$ as given by Eq. (\ref{rr1} ) ( with $E$ changed to $-E$ ) will
 give an  exact solution only if $s= \frac{ 1}{2}$. Most of the other properties in the last section 
continue to be valid in the DSG case also ( with the obvious replacement
of $E$ by $-E$ ) except now only $\bar{P}_n(E) $ and
$ \bar{Q}_n(E) $ exist while $\bar{R}_n(E) $ and $\bar{S}_n(E) $ as given by
Eq. (\ref{25}) do not exist ( Note that $P_n$ and $Q_n$ do not give QES solution if
$M$ is even integer). In particular, we want to emphasize that the
the norms of the polynomial sets $\{ P_n(E)\}, \{Q_n(E)\}, \{\bar{P}_n(E)\}$
and $ \{ \bar{Q}_n(E)\} $ are unchanged from those of the DSHG case. However the weight functions
 get interchanged . For example for $M=3$ the weight functions
for the DSG case are
\begin{eqnarray} 
w_0^{(p)} &=& \frac{ 1}{2} +\frac{ 2\zeta+1 }{2\sqrt{1+4\zeta^2}} \nonumber \\  
w_2^{(p)} &=& \frac{ 1}{2} -\frac{ 2\zeta+1 }{2\sqrt{1+4\zeta^2}} \nonumber \\  
w_1^{(q)} &=& 1
\end{eqnarray}
Thues even in this case, the weight function are {\it not} always positive.
Finally it is easily shown that the moments for the DSG and DSHG cases  are 
related by
\begin{equation}
\mu_n^{(p,q)}|_{DSG} = (-1)^n\mu^{(p,q)}_n|_{DSHG}
\end{equation}
so that for large n, to leading order, the DSG moments also have a pure power growth
rate as given by 
\begin{equation}
\mu_n\sim (-1)^n(M+\zeta)^{2n}.
\end{equation}

\subsection{New QES Potential }

We shall now point out a non-trivial application of the duality symmetry.
In particular, using this, we obtain an entirely new QES potential which has so far not been discussed in the literature \cite{book}.

Consider the following potential 

\begin{equation}
V(x) = \frac{ \mu^2 \left [ 8\sinh^4 \frac{ \mu x}{2}-4(\frac{ 5}
{\epsilon ^2}-1)\sinh^2 \frac{ \mu x}{2}+2 \left (\frac{ 1}{\epsilon ^4}
-\frac{ 1}{\epsilon ^2}-2 \right ) \right ]}{ 8 \left [1+\frac{ 1}{\epsilon ^2}+\sinh^2 
\frac{ \mu x}{2} \right ]^2}
\label{p1}
\end{equation}

which is obtained in the context of the stability analysis of the $\phi^6$
-kink solution in 1+1 dimensions \cite{lee}. As has been shown before, in this
case the Schro$\ddot{o}$dinger equation can be converted to Heun's equation
with four regular singular points. Further, for any $ \epsilon $, the ground
stat of the Schro$\ddot{o}$dinger equation corresponding to this
potential is given by $(\hbar =2m=1)$
\begin{equation}
\psi_0 = N_0 \left [\frac{  \epsilon ^2+1}{ 
\epsilon ^2+1+\epsilon ^2\sinh^2\frac{ \mu x}{2}} \right ] \left [ \frac{ \epsilon ^2\sinh^2\frac{\mu x}{2}} 
{ \epsilon^2+1+\epsilon^2\sinh^2 \frac{ \mu x}{2}} +\frac{ 1}{2} \right ]^{\frac{ 1}{2}},\ \ \  E_0 = 0
\end{equation}
Further for the special case of, $ \epsilon^2= \frac{ 1}{2}$, the second
excited
state is also analytically known and given by

\begin{equation}
\psi_2 = N_2 \left [ \frac{ 3}{3+\sinh^2\frac{ \mu x}{2}} \right ]\left [ \frac{ \sinh^2\frac{ 
\mu x}{2}}{3+\sinh^2 \frac{ \mu x}{2}} -\frac{ 1}{4} \right ],\ \ \  E_2 =\frac{ 3}{4}\mu^2 
\end{equation}
Thus for $ \epsilon^2=\frac{ 1}{2}$ this is a QES system. It may be noted
 that even though it is a QES system in one dimension,  the Hamiltonian can not
 be written in this case in terms of the quadratic generators of $Sl(2)$
\cite{jat}. Further, as  shown by us recently \cite{qes2}, in this case the
Bender-Dunne polynomials do not form an orthogonal set . We shall now
apply the duality transformation and obtain a new QES system
which has not been discussed before.

On considering the duality transformation $x\rightarrow i\theta $ in the Schro$\ddot{o}$dinger
equation $H\psi =E\psi$ corresponding to the potential (\ref{p1}), we find that
we have a new periodic potential 
  
\begin{equation}
\bar{V}(\theta ) =-\frac{ \mu^2 \left [ 8\sin^4 \frac{ \mu \theta }{2}+4(\frac{ 5}
{\epsilon ^2}-1)\sin^2 \frac{ \mu \theta }{2}+2 \left (\frac{ 1}{\epsilon ^4}
-\frac{ 1}{\epsilon ^2}-2 \right ) \right ]}{ 8 \left [1+\frac{ 1}{\epsilon ^2}-\sin^2 
\frac{ \mu \theta }{2} \right ]^2}
\label{pr2}
\end{equation}

Notice that the domain of $V(x)$ and $\bar{V}(\theta )$ are very different i.e. whereas 
$-\infty\le \mu x\le \infty,  \ \ 0\le\mu \theta \le 2\pi$. However, the
predictions of duality symmetry as given by Eq. (\ref{35}) are still valid.
In particular, using Eq. (\ref{35}) we predict that the potential $V(\theta )$ as given above is
a QES system in case $\epsilon ^2= \frac{ 1}{2}$ and it's ground and second
excited states energy eigenvalues are given by
\begin{eqnarray}
E_0 &=& - \frac{ 3}{4} \mu ^2,\ \ \  \ 
\psi_0= N_3 \left [\frac{ 3}{ 
3-\sin^2 \frac{ \mu \theta }{2}} \right ] \left [\frac{ 1}{4} +
\frac{ \sin^2 \frac{ \mu \theta }{2}}{3-\sin^2 \frac{ \mu
\theta }{2}} \right ]
 \nonumber \\  
E_2 &=& 0, \ \ \ 
 \psi_2 = N_4 \left [\frac{  \epsilon^2+1}{ 
\epsilon^2+1-\epsilon^2\sin^2 \frac{ \mu \theta }{2}} \right ] \left [ \frac{ 1}{2} -
\frac{ \epsilon ^2\sin^2 \frac{ \mu \theta }{2}}{ \epsilon^2+1-\epsilon^2\sin^2 \frac{ \mu
\theta }{2}} \right ]
\end{eqnarray}
In particular, we would like to emphasize that for the QES potential (\ref{pr2}),
the second excited state $\psi_2$ is known at all real values of $\epsilon$, 
while $\psi_0 $ is only known at  $ \epsilon^2 = \frac{1}{2}$.
One can solve the Schro$\ddot{o}$dinger 
equation explicitly for the potential (\ref{pr2}) and check that indeed these are the energy eigenstates of the system. 
In particular, note that both $\psi_0$ and $\psi_2$ 
are invariant under $\mu \theta \rightarrow \mu \theta +2\pi$ i.e.
$\psi(\mu \theta +2\pi) = \psi(\mu \theta )$. 
It is really remarkable that using duality we are able
to obtain a new QES system.
In this case  one can also obtain the three-term recursion relation 
for the Bender-Dunne polynomials  \cite{bd} and show that they do not form an
orthogonal set. This is related to the fact that the corresponding
Schro$\ddot{o}$dinger equation can be converted to Heun's equation with
four regular singular points.

 It may be worthwhile to look at all known QES
systems and see if the corresponding dual systems have already been discussed
in the literature or not. 
In this way one may discover some more QES systems.

\section{Discussions}
In this paper we have analyzed in some detail a QES system for which the
energy eigenstates for levels with odd as well as even number of nodes are known
for a given potential. 
 We have seen that in this case one obtains
two independent sets of orthogonal polynomials.
Further, we have also seen that the weight
functions in this case are not necessarily positive.
 It will be interesting to study
few other examples of the same type and enquire if in those cases too one
obtains two sets of orthogonal polynomials or not. Further, whether the weight functions
in such cases are always positive or not.
We have also compared the spectrum of two dual potentials in case
their domain of validity are quite different. We have seen in one case
that because of different boundary conditions, the QES spectrum of the dual
potentials is not the same. It will be interesting to study several such QES
dual potentials and see if some general conclusions can be obtained in these
cases.

Finally, it would be really interesting if one can discover some
 new QES potentials by making use of the duality symmetry.

\end{document}